\documentclass[onecolumn,aps,12pt]{revtex4}
\usepackage{amsmath,graphicx}
\usepackage{amsmath,graphicx}

\newcommand{\bea}{\begin{eqnarray}}
\newcommand{\eea}{\end{eqnarray}}

\newcommand{\e}{\varepsilon}
\newcommand{\de}{\Delta\varepsilon}
\newcommand{\LL}{L_\Lambda}

\newcommand{\bx}{\bar x}
\newcommand{\xb}{\bar x}
\newcommand{\by}{\bar y}
\newcommand{\yb}{\bar y}

\begin{document}


\title{Radiative corrections to radiative $\pi e 2$ decay}

\author{Yu.M.~Bystritsky}
\affiliation{Joint Institute for Nuclear Research, 141980 Dubna,
Russia}
\author{E.A.~Kuraev}
\affiliation{Joint Institute for Nuclear Research, 141980 Dubna,
Russia}
\author{E.P.~Velicheva}
\affiliation{Joint Institute for Nuclear Research, 141980 Dubna,
Russia}

\date{\today}

\begin{abstract}

    The lowest order radiative corrections (RC) to width and spectra of radiative $\pi e2$
    decay are calculated. We take into account virtual photon emission contribution as well
    as soft and hard real photon emission one. The result turns out to be consistent with the
    standard Drell-Yan picture for the width and spectra in the leading logarithmical
    approximation which permits us to generalize it to all orders of perturbation
    theory. Explicit expressions of nonleading contributions are obtained. The contribution
    of short distance is found to be in agreement with Standard
    Model predictions. It is presented as a general normalization factor. We check the validity of
    the Kinoshita-Lee-Nauenberg theorem about cancellation in the total width of the mass
    singularities at zero limit of electron mass. We discuss the results of the previous
    papers devoted to this problem. The Dalitz plot distribution is illustrated
    numerically.

\end{abstract}


\maketitle


\section{Introduction}
\label{sect1}

The process of radiative negative pion decay
\begin{gather}
\pi^-(p) \to e^-(r) + \bar \nu_e(q) + \gamma(k)
\end{gather}

\noindent attracts a lot of attention both experimentally and
theoretically \cite{Bryman:et}, \cite{Pocanic:2003tp},
\cite{Bardin:ia}, \cite{Nikitin}, \cite{Komachenko:1992hr},
\cite{Belyaev:1991gs}, \cite{Poblaguev:2003ib}. The main reason is
a unique possibility to extract the so-called structure-dependent
part $M_{SD}$ of the matrix element
\bea
    M = M_{IB} + M_{SD},
\eea

\noindent which can be described in terms of vector and
axial-vector formfactors of the pion and the hints of a "new
physics" including a possible revealing of tensor forces (which
are absent in the traditional Standard Model).

Due to the numerical smallness of $M_{SD}$ compared to the inner
bremsstrahlung part $M_{IB}$, the problem of taking into account
radiative corrections (RC) becomes essential. The lowest order RC
for a special experimental set-up was calculated in 1991 in
\cite{Nikitin}, where, unfortunately, the contribution to RC from
emission of an additional hard photon was not considered. This is
the motivation of our paper.

Our paper is organized as follows. In part
\ref{VirtualAndSoftContributions}, we calculate the  contributions
to RC from emission of virtual photons (at 1 loop level) and the
ones arising from additional soft photon emission. For
definiteness, we consider RC to the QED-part of the matrix element
$M_{IB}$
\bea
    M_{IB} = -i A~\bar u_e(r)
    \left[
        \frac{(\e^* r)}{(k r)}-\frac{(\e^* p)}{(k p)}
        +\frac{\hat\e^* \hat k}{2(k r)}
    \right]
    (1+\gamma_5)v_\nu(q),
    \qquad
    A=e\frac{G_F}{\sqrt{2}} V_{ud} f_\pi m,
    \label{BornAmplitudeIBPartWithADefined}
\eea

\noindent where $e=\sqrt{4\pi\alpha}$, $G_F = 1.17\cdot 10^{-5}~
GeV^{-2}$ is the Fermi coupling constant, $\e$ is the photon
polarization vector, $f_\pi = 131~MeV$ is the pion decay constant,
$V_{ud}$ is the element of the Cabbibo-Kobayashi-Maskawa quark
mixing matrix, and $M$ and $m$ are masses of pion and electron.
The explicit form of the matrix element including $M_{SD}$ is
given in Appendix \ref{AppendixA}.

In part \ref{HardContributions}, we consider the process of double
radiative pion decay and extract the leading contribution
proportional to "large" logarithm $\ln M^2/m^2$ which arises from
the kinematics of emission of one of the hard photons collinearly
to electron momentum. We arrive at the result which can be
obtained by applying the quasireal electron method \cite{BFK}.

In conclusion, we combine the leading contributions and find that
the result can be expressed in terms of the electron structure
function \cite{Kuraev:hb}. Really, the lowest order RC in the
leading logarithmical approximation (LLA) (i.e., keeping terms
$\left(\frac{\alpha}{\pi}\ln \frac{M^2}{m^2}\right)^n$)  turns out
to reproduce two first order contribution of the electron
structure function obeying the evolution equation of
renormalization group (RG). This fact permits us to generalize our
result to include higher orders of perturbation theory (PT)
contributions to the electron structure function in the leading
approximation. In conclusion, we also argue that our consideration
can be generalized to the whole matrix element.

As for the next-to-leading contributions, we put them in the form
of the so-called $K$-factor which collects all the non-enhanced
(by large logarithm) contributions. Part of them, arising from
virtual and real soft photon emission, is given analytically. The
other part, arising from emission of additional hard photon in
noncollinear kinematics is presented in Appendix \ref{AppendixC}
in terms of 3-fold convergent integrals.

In Appendix \ref{AppendixB}, we give the simplified form of RC
(the lowest order ones) and make an estimation of the omitted
terms which determine the accuracy of the simplified RC.

Appendix \ref{AppendixD} contains the list of 1-loop integrals
used in calculation of 1-loop Feynman integrals.

In tables \ref{BornIBDalitzTable}, \ref{IBRCDalitzTable},
\ref{KDalitzTable} the result of numerical calculation of
contributions of RC to Dalitz-plot distribution in Born level,
leading and non-leading approximations  are given.

\section{Virtual and soft real photon emission RC}
\label{VirtualAndSoftContributions}

A rather detailed calculation of the lowest order RC was carried
out in \cite{Nikitin}. Nonclear manipulations with a soft photon
emission contribution used in \cite{Nikitin} results in a wrong
form of the dependence of RC on the photon detection threshold
$\de$ which contradicts the general theorem \cite{Yennie:ad}.
Another reason for our revision of \cite{Nikitin} is the mixing of
(on-mass shell and dimensional) regularization schemes in it. We
use here the unrenormalized theory with the ultraviolet cut-off
parameter $\Lambda$.

Following \cite{Nikitin}, we first consider RC to the "largest"
contribution -- inner bremsstrahlung $M_{IB}$. As well we consider
the central part of Dalitz-plot and omit (if possible) $\beta
\equiv \left(\frac{m}{M}\right)^2 = 1.34\cdot 10^{-5}$, that is
$\beta \rightarrow 0$.

We distinguish the contribution to RC from emission of virtual and
soft additional photons ($\Delta_V$ and $\Delta_S$, respectively):
\bea
    \sum_{spin~states} \left.\left|M_{IB}\right|^2\right|_{virt+soft} =
    \sum \left|M_{IB}\right|^2
    \left(1+\frac{\alpha}{\pi}\left(\Delta_V+\Delta_S\right)\right),
\eea

\noindent where
\bea
    \sum \left|M_{IB}\right|^2 =
    8A^2 \frac{(1-y)(1+(1-x)^2)}{x^2(x+y-1)}
    \equiv 8 A^2 B(x,y), \label{IB}
\eea

\noindent (here $x=\frac{2(k p)}{M^2}$, $y=\frac{2(r p)}{M^2}$,
$z=\frac{2(k r)}{M^2}$, $M$ is the pion mass).

Soft photon emission RC have a standard form (see, for example
\cite{Bytev:2002nx}, formula (16))
\bea
    \frac{\alpha}{\pi} \Delta_S &=&
    -\frac{\alpha}{2\pi^2} \int \frac{d^3k_1}{2\omega_1}
    \left.
    \left(\frac{p}{(k_1 p)}-\frac{r}{(k_1 r)}\right)^2
    \right|_{\omega_1=\sqrt{{\vec k_1}^2+\lambda^2} < \de \ll y M}= \nonumber \\
    &=&
    \frac{\alpha}{\pi}\left[
    \left(L_e-2\right) \ln\left(\frac{2 \de}{\lambda}\right) +
    \frac{1}{2}L_e - \frac{1}{4}L_e^2 + 1 - \xi_2
    \right]
    \left( 1 + O(\beta) \right),
\eea

\noindent where $k_1$ the additional soft photon momentum,
$L_e=\ln\frac{y^2}{\beta}$ and $\lambda$ is the "photon mass",
$\xi_2=\frac{\pi^2}{6}$. This result agrees with the general
analysis of infrared behavior given in \cite{Yennie:ad}.

Now let us consider the calculation of the virtual photon emission
corrections.
First, we use the minimal form for introduction of the
electromagnetic field through the generalization of the derivative
$p_\mu \rightarrow p_\mu - i e A_\mu$.

The sum of contributions (all particles except antineutrino can be
on or off mass shell) leads to:
\bea
    \frac{\hat\e(\hat r + \hat k + m)\hat p}{(r+k)^2 -
    m^2}v_\nu(q) - \hat \e v_\nu(q) =
    m \cdot \frac{\hat\e(\hat r + \hat k + m)}{(r+k)^2 -
    m^2}v_\nu(q),
\eea

\begin{figure}
\includegraphics[scale=1]{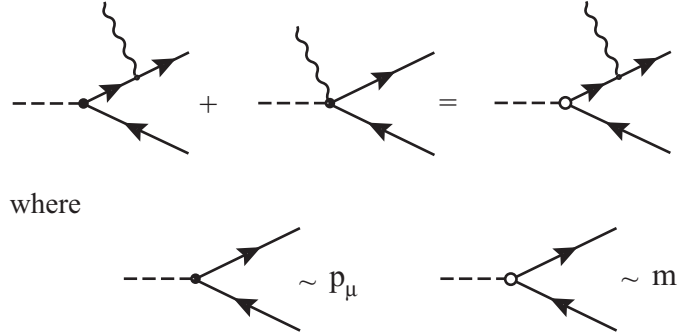}
\caption{Effective vertex.} \label{EffectiveVertex}
\end{figure}

\noindent $p=r+k+q$. Thus, we can extract the contact photon
emission vertex and obtain the effective vertex $\sim m$, as it is
shown in Fig. \ref{EffectiveVertex}.

\begin{figure}
\includegraphics[scale=1]{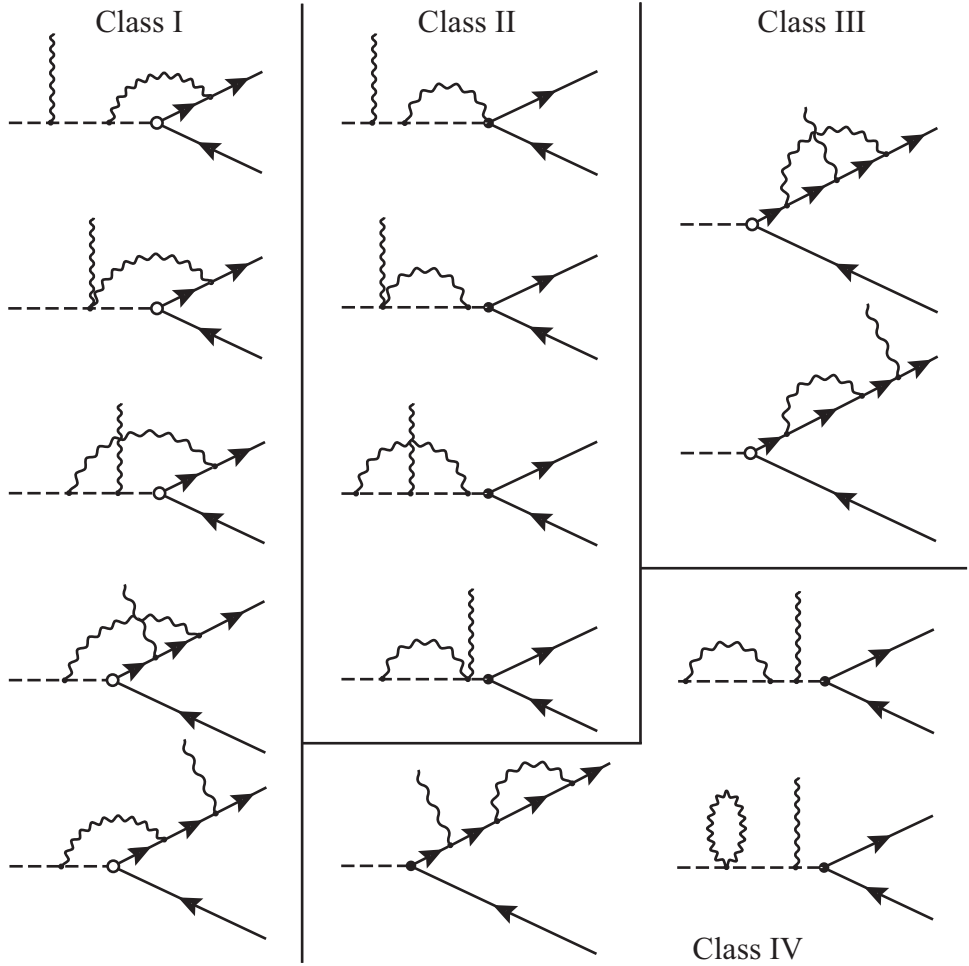}
\caption{Feynman diagrams of virtual contributions.}
\label{VirtualCorrectionsInitialFDPicture}
\end{figure}

In terms of this new effective vertex we can write down the 1-loop
Feynman diagrams (FD) of virtual photon emission RC
($\frac{\alpha}{\pi}\Delta_V$) which are shown in Fig.
\ref{VirtualCorrectionsInitialFDPicture} (see \cite{Nikitin}). We
should notice that these diagrams can be separated out into four
classes. The contributions of each of these classes are gauge
invariant.

Three first classes were considered in \cite{Nikitin}, where their
gauge invariance and in particular zero contribution of class II
were strictly shown. The last statement directly follows from the
gauge-invariance.

The contribution of class III contains the regularized mass and
vertex operators. The relevant matrix element has the form:
\bea
    M_{III}=-i A \frac{\alpha}{2\pi M} \bar u_e(r)
    \left[
        \left(
              \hat \e^* - \hat k \frac{(\e^* r)}{(k r)}
        \right) B_1
        +
        \frac{\hat k \hat \e^*}{M} B_2
    \right]
    (1+\gamma_5) v_\nu(q),
\eea

\noindent where
\bea
    B_1 &=& \frac{1}{\sqrt{\beta}} \frac{1}{2a}
        \left( 1-\frac{y'}{a}l_S \right), \nonumber \\
    B_2 &=& \frac{1}{\beta}
        \left[
            \frac{1}{y'}n + \frac{1}{2a}-
            \frac{2y'^2+3y'+2}{2 y' a^2}l_S
        \right], \nonumber \\
    n &=& \frac{1}{y'}
        \left[
            Li_2(1)-Li_2(1+y')-i\pi\ln(1+y')
        \right], \nonumber
\eea

\noindent here $l_S=\ln \frac{z}{\beta}-i\pi$,~~$a=1+y'$,
~$y'=\frac{2(k r)}{m^2}$. We see that $M_{III}$ is explicitly
gauge-invariant and is free from infrared singularities. At the
realistic limit $y \sim x \sim z \gg \beta$ (rather far from the
boundaries of Dalitz-plot) we obtain for contribution to the
matrix element square (structure $\sim B_1$ gives a zero
contribution in the limit $\beta \to 0$) (in agreement with
\cite{Nikitin}):
\bea
    \Delta\left|M_{III}\right|^2 =
    2 Re\left\{ \left(M_{IB}\right)^+ M_{III}\right\} =
    -\frac{8 \alpha}{\pi} A^2 \frac{1-y}{x z}
    \left(\frac{1}{2}-\ln\left(\frac{z}{\beta}\right)\right).
\eea

\noindent As we work within the unrenormalized theory, we should
consider class IV (not considered in \cite{Nikitin}) which
concerns the contribution of counter-terms due to renormalization
of the pion and electron wave functions (see \cite{Bytev:2002nx},
formula (17))
\bea
    \Delta\left|M_{IV}\right|^2 =
    \sum \left|M_{IB}\right|^2
    \cdot \frac{\alpha}{2\pi}
    \left\{
        \left[-\frac{1}{2}\LL + \frac{3}{2} \ln \beta +
                \ln\frac{M^2}{\lambda^2} - \frac{9}{4}
        \right]
        +
        \left[\LL + \ln\frac{M^2}{\lambda^2} - \frac{3}{4}
        \right]
    \right\},
\eea

\noindent where $\LL=\ln\frac{\Lambda^2}{M^2}$ and $\Lambda$ is
the ultraviolet cut-off parameter to be specified later. The first
term in the brackets in the r.h.s. is the electron wave function
renormalization constant and the second one corresponds to the
pion wave function renormalization.

Let now consider the contributions of FDs of class I. In
\cite{Nikitin} the explicit gauge-invariance of the sum was
demonstrated: $k_\mu \left( T_1^\mu + T_2^\mu + T_3^\mu + T_4^\mu
+ T_5^\mu \right) = 0$, where $T_i^\mu$ is the contribution of the
corresponding FD of Class I (see in Fig.
\ref{VirtualCorrectionsInitialFDPicture}).

The contribution to the matrix element square can be written as:
\bea
    \Delta\left|M_{I}\right|^2 =
    2 Re\left\{ \left(M_{IB}\right)^+ M_{I}\right\} =
    \frac{\alpha}{2\pi} A^2 Re \left\{\int \frac{d^4k_1}{i \pi^2}
    \left[ A_1 + A_2 + A_3 + A_4 + A_5 \right] \right\},
\eea

\noindent where
\bea
    A_1 &=& \frac{2}{x}\cdot\frac{1}{(0)(1')(2)}
            Sp~[
                \hat r (2(\hat p - \hat k) - \hat k_1)(\hat r - \hat k_1)
                (1+\gamma_5)
                \hat q \left((Q p) + \frac{\hat k \hat
                p}{z}\right)
            ]\frac{1}{M^2}, \nonumber \\
    A_2 &=& \frac{2}{(0)(1')(2)}
            Sp~[
                \hat r \gamma_\lambda (\hat r - \hat k_1)
                (1+\gamma_5)
                \hat q \left(Q^\lambda + \frac{\hat k \gamma^\lambda}{z}\right)
            ], \nonumber \\
    A_3 &=& -\frac{2}{(0)(1)(1')(2)}
            Sp~[
                \hat r (2\hat p - \hat k_1)(\hat r - \hat k_1)
                (1+\gamma_5)
                \hat q \left((Q (p-k_1)) + \frac{\hat k (\hat p-\hat k_1)}{z}\right)
            ], \nonumber \\
    A_4 &=& -\frac{1}{(0)(1)(2)(2')}
            Sp~[
                \hat r (2\hat p - \hat k_1)(\hat r - \hat
                k_1)\gamma_\lambda (\hat r + \hat k - \hat k_1)
                (1+\gamma_5)
                \hat q \left(Q^\lambda + \frac{\hat k \gamma^\lambda}{z}\right)
            ], \nonumber \\
    A_5 &=& -\frac{1}{z}\frac{1}{(0)(1)(2')}
            Sp~[
                \hat r \gamma_\lambda (\hat r + \hat k) (2\hat p - \hat k_1) (\hat r + \hat k - \hat k_1)
                (1+\gamma_5)
                \hat q \left(Q^\lambda + \frac{\hat k \gamma^\lambda}{z}\right)
            ]\frac{1}{M^2}. \nonumber
\eea

\noindent where $k_1$ virtual photon momentum,
$Q_\mu=-\frac{p_\mu}{(kp)}+\frac{r_\mu}{(kr)}=\frac{2}{M^2}\left(\frac{r_\mu}{z}-\frac{p_\mu}{x}\right)$.
Here the denominators of the pion and electron Green functions
$(i)$ are listed in Appendix \ref{AppendixD}. We omitted $m$
everywhere in the numerators. The ultraviolet divergences are
present in $A_1$ and $A_5$.

Using the set of vector and scalar integrals listed in Appendix
\ref{AppendixD} and adding the soft photon emission corrections we
obtain
\bea
    &&1 + \frac{\alpha}{\pi}\left(\Delta_V+\Delta_S\right)=
    \nonumber\\
    &&\qquad =
    S_W \left\{
        1 + \frac{\alpha}{\pi}
        \left[
            \left(L_e-1\right)
            \left(\ln\Delta + \frac{3}{4}\right)
            - \ln \Delta
            + \frac{1}{B'(x,y)}\cdot R(x,y)
        \right]
        \right\},
\eea

\noindent where $S_W = 1+\frac{3}{4}\frac{\alpha}{\pi} \LL$,
$\Delta \equiv \frac{2\de}{y M}$. $R(x,y)$ looks like
\bea
    R(x,y)&=& 4 a b \by \left(
                    - 5\ln y + \frac{3}{2}
                    -4\ln y \ln\frac{2}{y}
                \right)-\nonumber \\
    &-&8 b \ln y \left(
                    -a \by (1+\ln 4)
                    +((x-2)^2-y(a+2)) \ln x
                    +x(1+y^2) \ln z
                \right)+\nonumber \\
    &+&2 b \ln^2 y \left(
                        -16 \by
                        +x(16-7x+7y(x-2))
                    \right)-\nonumber \\
    &-&\xi_2\cdot 8 b \by (6+2x^2-x(5+y))+\nonumber\\
    &+&4 x \by z b \left(
                        Li_2\left(\frac{y-1}{y}\right)-
                        Li_2(1-y)
                    \right)
    +8 a b \by Li_2(x)+\nonumber \\
    &+&8 a b \by \ln x \ln(1-x) - 8 x \bx \by z (x+y-2) \ln
    z-\nonumber\\
    &-& 4 x \by z \ln^2 z (-\bx + y(x-2)+y^2) - 4 x \by b,
    \label{K_soft}
\\
    B'(x,y) &=& 8 (1-y)(1+(1-x)^2)(x+y-2)^2 = 8 a b \by
\label{B_prime}, \eea \noindent where $a = 1+(1-x)^2$,
$b=(x+y-2)^2$, $\bx= 1-x$, $\by= 1-y$, $z=x+y-1$. Here we do not
have a complete agreement with the result of \cite{Nikitin}. Note
that the sum $\Delta_S+\Delta_V$ does not depend on "photon mass"
$\lambda$, and, besides, the coefficient at large logarithm $L_e$
agrees with RG predictions.

\section{Emission of additional hard photon}
\label{HardContributions}

\begin{figure}
\includegraphics[scale=1]{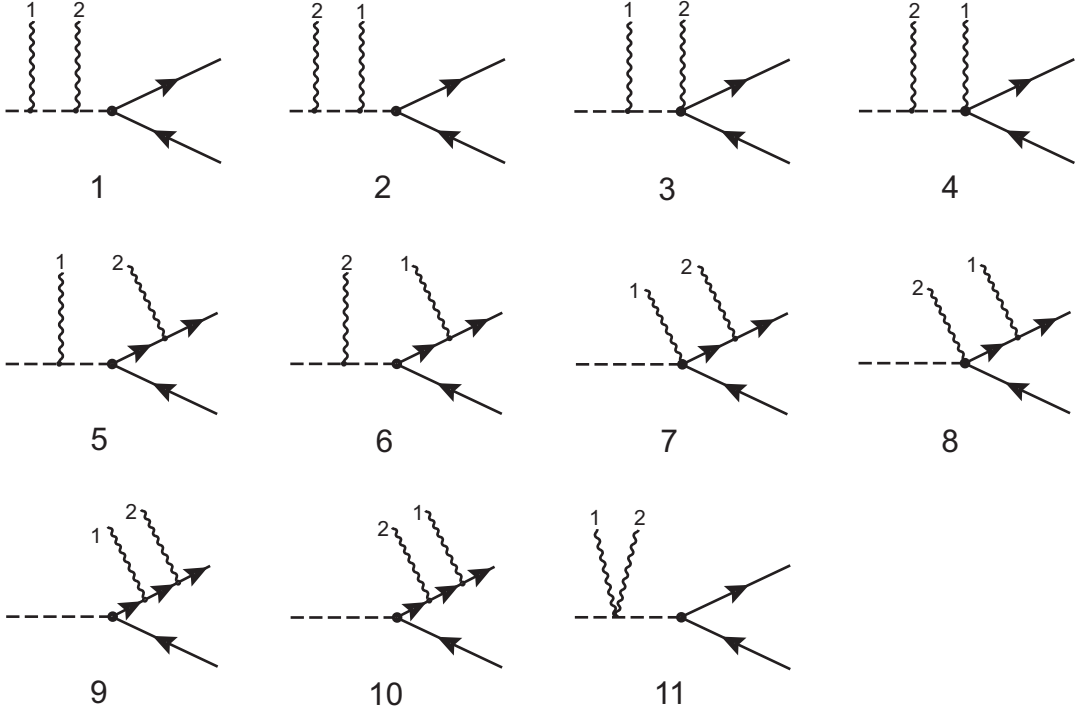}
\caption{Feynman diagrams, describing the double radiative pion
decay.} \label{HardPhotonEmissionFDPicture}
\end{figure}

Emission of additional hard photon (not considered in the previous
papers concerning RC to the $\pi \to e \nu \gamma$ decay), i.e.,
the process of double photon emission in the $\pi e 2$ decay:
$\pi^-(p) \rightarrow e^-(r)+\bar\nu_e(q)+\gamma(k_1)+\gamma(k_2)$
is described by 11 FD drawn in Fig.
\ref{HardPhotonEmissionFDPicture}.

In collinear kinematics (when one of the real photons is emitted
close to the electron emission direction) the relevant
contribution to the differential width contains large logarithms.

The matrix element of the double radiative pion decay has the form
\bea
    M^{\pi\rightarrow e \nu \gamma\gamma} &=& i A\sqrt{4\pi \alpha}~
    \left[\bar u_e(r)O_{\mu\nu}(1+\gamma_5)v_\nu(q)\right]
    \cdot \e_1^\mu(k_1) \e_2^\nu(k_2),
\eea

\noindent where $\e_{1,2}$ are the photons polarization vectors
which obey the Lorentz condition: $(k_1~\e_1(k_1)) =
(k_2~\e_2(k_2)) = 0$. The tensor $O^{\mu\nu}$ has the form
\bea
    O^{\mu\nu} &=&
        \left(
            - 2g^{\mu\nu}
            + \frac{2 p^\mu \cdot 2(p-k_1)^\nu}{-2(k_1 p)}
            + \frac{2 p^\nu \cdot 2(p-k_2)^\mu}{-2(k_2 p)}
        \right)
        \frac{1}{(p-k_1-k_2)^2-M^2}
        + \nonumber \\
    &+&
        \frac{2 p^\mu}{-2(k_1 p)}
        \frac{\gamma^\nu (\hat r + \hat k_2 + m)}{2(k_2 r)}
        +
        \frac{2 p^\nu}{-2(k_2 p)}
        \frac{\gamma^\mu (\hat r + \hat k_1 + m)}{2(k_1 r)}
        + \nonumber \\
    &+&
        \left[
            \gamma^\nu \frac{\hat r + \hat k_2 + m}{2(k_2 r)}
            \gamma^\mu
            +
            \gamma^\mu \frac{\hat r + \hat k_1 + m}{2(k_1 r)}
            \gamma^\nu
        \right]
        \frac{\hat r + \hat k_1 + \hat k_2 +
        m}{(r+k_1+k_2)^2-m^2}.
\eea

\noindent One can be convinced in the explicit fulfillment of the
requirements of Bose-symmetry and gauge-invariance
$M^{\pi\rightarrow e \nu \gamma\gamma}(\e_1\rightarrow k_1) =
M^{\pi\rightarrow e \nu \gamma\gamma}(\e_2\rightarrow k_2) = 0$.
The expression for $\sum \left|M^{\pi\rightarrow e \nu
\gamma\gamma}\right|^2 \sim Sp\left[(\hat r+m)O^{\mu\nu}\hat q~
O_{\mu\nu} \right]$ is rather cumbersome.

The contribution to the differential width has the form
\bea
    &&d \Gamma^{hard}_{IB}=\frac{1}{2M}
    \sum \left|M^{\pi\rightarrow e \nu \gamma\gamma}\right|^2
    d \Phi_4,
\eea
\bea
    &&d \Phi_4 = \frac{(2\pi)^4}{(2\pi)^{12}}
    \frac{d^3q}{2q_0} \frac{d^3r}{2r_0}
    \frac{d^3k_1}{2\omega_1}\frac{d^3k_2}{2\omega_2}~
    \delta^4(p-q-r-k_1-k_2).
\eea

\noindent We do not take into account the identity of photons: we
believe the photon with momentum $k_1$ to be a measurable one with
$\frac{2(k_1p)}{M^2}=x$ and the photon with momentum $k_2$ to be a
background one with $\frac{2(k_2p)}{M^2}>\Delta$ and most general
kinematics.

It is convenient to consider separately the collinear kinematics
of emission of one of the photons, i.e., the case when the angle
of emission of one of the final photon direction of motion is
rather small $\theta_i=(\widehat{\vec r,\vec k_i}) < \theta_0 \ll
1$, $\theta_0^2 \gg \beta$. (Note that double collinear kinematics
is excluded since the invariants $x'_{1,2}=\frac{2(k_{1,2}
r)}{M^2}$ cannot be small simultaneously).

The contribution of these collinear kinematics to the differential
width contains "large logarithms" $L_e$. To extract the relevant
contribution we can use the quasireal electron method \cite{BFK}.
For this aim let us arrange the integration over phase volume in
the following way:
\bea
    \left|M\right|^2 d \Phi_4 =
    \left|M_C\right|^2 d \Phi_4^C+
    \left|M_C\right|^2 (d \Phi_4-d \Phi_4^C)+
    \left(\left|M\right|^2-\left|M_C\right|^2\right)d \Phi_4,
    \label{dPhi4Spliting}
\eea

\noindent with
\bea
    (4\pi\alpha)^{-2} \sum \left|M_C\right|^2=
    \left[
        \frac{8}{x'_2}
        \frac{y^2+(y+x_2)^2}{x_2(x_2+y)}
        -
        \frac{16\beta}{{x'_2}^2}
    \right]
    \cdot
    \frac{(1+(1-x_1)^2)(1-x_2-y)}{x_1^2(x_1+x_2+y-1)},
\eea
\bea
    (2\pi)^8 M^{-4}d \Phi_4 &=&
    M^{-4}\frac{d^3r~d^3k_1~d^3k_2}{2r_0~2\omega_1~2\omega_2}~
    \delta\left(\left(p-r-k_1-k_2\right)^2\right) = \nonumber \\
    &=&
    \frac{\pi^2}{2^6}~x_1x_2y~d x_1 d x_2 dy~dC_1~d\Omega_2~ \times
    \nonumber\\
    &\times&
    \delta\left(
        1-x_1-x_2-y
        +\frac{x_1x_2}{2}(1-C_{12})
        +\frac{x_1 y}{2}(1-C_{1})
        +\frac{x_2 y}{2}(1-C_{2})
    \right), \label{PhaseVolume}\\
    (2\pi)^8 M^{-4}d \Phi_4^C &=&
    \frac{\pi^2}{2^5}~\frac{x_2 y}{y+x_2} d x_1 d x_2 dy~d\Omega_2^C,
\eea

\noindent where $x_{1,2}=\frac{2(k_{1,2}
p)}{M^2}$,~$x'_{1,2}=\frac{2(k_{1,2} r)}{M^2}$,~
$C_{12}=\cos(\widehat{\vec k_1,\vec k_2})$,~
$C_{1,2}=\cos(\widehat{\vec k_{1,2},\vec r})$,~$d\Omega_2$ is the
angular phase volume of the photon with 4-momentum $k_2$.

Note that integration over the angular phase volume $d\Omega_2$ in
$d \Phi_4^C$ (see (\ref{dPhi4Spliting})) is restricted by $\int
d\Omega_2^C=\int\limits_0^{2\pi} d\phi
\int\limits_{1-\theta_0^2/2}^{1} dC_2$ and in the second term in
(\ref{dPhi4Spliting}) $d \Phi_4-d \Phi_4^C$ can be replaced by $d
\Phi_4$ with $\int d\Omega_2=\int\limits_0^{2\pi} d\phi
\int\limits_{-1}^{1-\theta_0^2/2} dC_2$.
The second and the third terms in (\ref{dPhi4Spliting}) do not
contain collinear singularities, i.e., are finite in the limit
$\beta\rightarrow 0$.

The contribution of hard photon emission can be written out in the
form
\bea
    \frac{d\Gamma^{hard}_{IB}}{d x d y}&=&
    A^2 \frac{\alpha}{2\pi} \int\limits_y^1
    \frac{dt}{t} B(x,t)
    \left[
        P_\theta^{(1)}\left(\frac{y}{t}\right)
        \left(
            L_e-1
            +
            \ln\frac{\theta_0^2}{4}
        \right)
        +
            1-\frac{y}{t}
    \right]
    +\nonumber \\
    &+&
    \frac{\alpha}{\pi}
    K^{hard}_{IB}(x,y,\theta_0,\Delta),
\eea

\noindent where the function $B(x,t)$ is defined in (\ref{IB}),
\bea
    P_\theta^{(1)}(z)=\frac{1+z^2}{1-z}
    \theta\left(1-z-\Delta\right),
\eea

\noindent and $K^{hard}_{IB}(x,y,\theta_0,\Delta)$ represents the
contribution of two last terms in (\ref{dPhi4Spliting}).
Now it is convenient to introduce the following quantity:
\bea
    K^{h}_{IB}(x,y,\Delta) = K^{hard}_{IB}(x,y,\theta_0,\Delta)+
    \int\limits_y^1
    \frac{dt}{t} B(x,t)
    \left[
        P_\theta^{(1)}\left(\frac{y}{t}\right)
            \ln\frac{\theta_0^2}{4}
        +
            1-\frac{y}{t}
    \right], \label{Kh_factor}
\eea

\noindent which does not already depend on the parameter
$\theta_0$. The explicit view of $K^{h}_{IB}(x,y,\Delta)$ is given
in Appendix \ref{AppendixC}.

In the total sum of virtual, soft, and hard photon emission
contributions all the auxiliary parameters -- "photon mass"
$\lambda$ and $\Delta$ -- are cancelled out. The resulting
expression for the differential width with RC up to any order of
QED PT with the leading logarithm $\frac{\alpha}{\pi}L_e\sim 1$
and the next-to-leading accuracy have the form
\bea
    \frac{d\Gamma^{RC}_{IB}}{d x d y}=
    A^2 S_W \int\limits_y^1
    \frac{dt}{t} B(x,t)
    D\left(\frac{y}{t}\right)
    \left(
        1+\frac{\alpha}{2\pi}K_{IB}(x,y)
    \right), \label{RCResult}
\eea
where $D(z)$ is the well-known electron structure function, which
has the form
\bea
    D(z)=\delta(1-z) +
    \frac{\alpha}{2\pi}\left(L_e-1\right) P^{(1)}(z) +
    \frac{1}{2!}\left(\frac{\alpha}{2\pi}\right)^2\left(L_e-1\right)^2
    P^{(2)}(z)+
    \cdots,
\eea
where
\bea
    P^{(1)}=\lim_{\Delta\rightarrow0}
    \left[
        \frac{1+z^2}{1-z}\theta(1-z-\Delta)
        +
        \delta(1-z)\left(2\ln\Delta+\frac{3}{2}\right)
    \right]
    =
    \left(
        \frac{1+z^2}{1-z}
    \right)_+,
\eea
\bea
    P^{(n)}(z)=\int\limits_z^1\frac{dt}{t}
    P^{(1)}(z)P^{(n-1)}\left(\frac{z}{t}\right).
\eea
The function $K_{IB}(x,y)$ is the so-called K-factor which here
has the form
\bea
    K_{IB}(x,y)=2\frac{R(x,y)}{B'(x,y)} -2\ln\Delta + K^{h}_{IB}(x,y,\Delta),
    \label{K_IB}
\eea

\noindent where $R(x,y)$ and $B'(x,y)$ was defined in
(\ref{K_soft}) and (\ref{B_prime}). The quantity
$K^{h}_{IB}(x,y,\Delta)-2\ln\Delta$ is finite at $\Delta\to0$ and
its explicit expression is given in Appendix~\ref{AppendixC}.

\section{Discussion}

It is easy to see that in the total decay width all the dependence
on $\beta$ disappears in accordance with the
Kinoshita-Lee-Nauenberg (KLN) theorem \cite{KLN}. Really, we
obtain integrating over $y$:
\bea
    \int\limits_0^1 dy \int\limits_y^1\frac{dt}{t}
    D\left(\frac{y}{t}\right) \cdot f(t) =
    \int\limits_0^1 dt~f(t) \int\limits_0^t \frac{dy}{t}
    D\left(\frac{y}{t}\right) =
    \int\limits_0^1 dt~f(t) \int\limits_0^1 dz
    D\left(z\right) =
    \int\limits_0^1 dt f(t).
\eea

Now let us discuss the dependence on the ultraviolet cut-off
$\Lambda$. It was shown in a series of remarkable papers by
A.~Sirlin \cite{Sirlin:cs} that the Standard Model provides
$\Lambda=M_W$. Another important moment (not considered here) is
the evolution with respect to ultraviolet scale of virtual photon
momenta from hadron scale ($m_\rho$) up to $M_Z$ \cite{Sirlin:cs}.
It results in effective replacement $S_W \rightarrow
S_{EW}=1+\frac{\alpha}{\pi}\ln\frac{M_Z^2}{m_\rho^2}\approx1.0232.$
So all the QED corrections to the total width are small $\sim
O\left(\frac{\alpha}{\pi}\right)$, but the electroweak ones rather
large: $\Gamma \approx \Gamma_0 \cdot S_{EW}$. The factor $S_{EW}$
can be absorbed by pion life-time constant $f_\pi\sqrt{S_{EW}} \to
f_\pi^{exp}$ \cite{Holstein:ua}. Thus, we replace $A$, defined in
(\ref{BornAmplitudeIBPartWithADefined}), with $A_{exp.}$
\bea A=e\frac{G_F}{\sqrt{2}} V_{ud} f_\pi m \quad \to\quad
A_{exp.}=e\frac{G_F}{\sqrt{2}} V_{ud} f_\pi^{exp} m.
\label{AReplacement} \eea
We also note here that according to \cite{Holstein:ua} we must use
the redefined constant $A_{exp.}$ in the Born approximation. That
is why we use $A_{exp.}$ in Appendix \ref{AppendixA}.

In our explicit calculations we considered RC to the inner
bremsstrahlung part of the matrix element. Let us now argue that
in the integrand of the r.h.s. of (\ref{RCResult}) one can replace
$B(x,y)$ by the total value including the structure dependent
contribution $\Phi_{tot}^{(0)}(x,y)$ which is defined in Appendix
\ref{AppendixA} in (\ref{BornDecayWidth}). This fact can be proved
in the leading logarithmic approximation. One of contributions
arising from additional hard photon emission close to the electron
direction can be obtained by applying the quasireal electron
method \cite{BFK} and has the form (\ref{RCResult}) with $P^{(1)}
\rightarrow P^{(1)}_\theta$. The KLN theorem in a unique form
provides the soft photon emission and virtual RC to be of a form
with complete kernel $P(z)$ in the structure function $D(z)$ in
(\ref{RCResult}). So our result reads
%
%
\bea
    \frac{d\Gamma_{RC}}{d x d y}=
    A_{exp.}^2 \int\limits_y^1
    \frac{dt}{t} \Phi_{tot}^{(0)}(x,t)
    D\left(\frac{y}{t}\right)
    \left(
        1+\frac{\alpha}{2\pi}K(x,y)
    \right). \label{RCResultTotal}
\eea

We also calculate the RC to the SD part in leading logarithmical
approximation. Formula (\ref{RCResultTotal}) can be trusted in the
region where the IB part dominates. In the region where IB $\le$
SD, which is suitable for SD measurement, both IB and SD are
small; so the question about non-leading contributions becomes
academical. Thus, we suggest here that K-factor is the same order
of magnitude as for the IB part (i.e. $K(x,y)=K_{IB}(x,y)$), where
it can be calculated in the model-independent way. Its numerical
value is given in Table \ref{KDalitzTable}.



We underline that the explicit dependence on "large logarithm"
$L_e=\ln\frac{y^2}{\beta}$ is present in the Dalitz-plot
distribution.

Now let us discuss the results obtained in some previous papers
devoted to RC in the radiative pion decay.

In \cite{Nikitin}, the hard photon emission was not considered
which led to violation of the KLN-theorem.

In \cite{Komachenko:1992hr}, \cite{Belyaev:1991gs}, the main
attention was paid to possible sources of tensor forces. As for
real QED+EW corrections depicted in formula (13) in
\cite{Komachenko:1992hr}: $\frac{\delta\Gamma}{\Gamma}\approx 0.7
\frac{\alpha}{\pi}\sim 0.2 \%$ the QED leading RC was presumably
omitted.


\section{Acknowledgements}

We are grateful to RFFI grant 03-02 N 17077 for support. One of us
(E.K.) is grateful to INTAS grants 97-30494 and 00366. We are also
grateful to M. V. Chizhov for discussions and pointing out
reference \cite{Poblaguev:2003ib}. Besides, we are grateful to
Paul Scherrer Institute (Villigen, Switzerland) for warm
hospitality where the final part of this work was completed.

\appendix

\section{Born amplitude}
\label{AppendixA}

The radiative pion decay matrix element in the Born approximation
has the form \cite{Poblaguev:2003ib}
\bea
    M = M_{IB} + M_{SD} + M_T,
    \label{BornAmplitude}
\eea

\noindent where
\bea
    M_{IB} &=& -i A_{exp.}~\bar u_e(r)
    \left[
        \frac{(\e^* r)}{(k r)}-\frac{(\e^* p)}{(k p)}
        +\frac{\hat\e^* \hat k}{2(k r)}
    \right]
    (1+\gamma_5)v_\nu(q), \\
    M_{SD} &=& \frac{A_{exp.}}{m M f_\pi^{exp.}}~
    \bar u_e(r)\gamma_\mu (1+\gamma_5)v_\nu(q)~e_\nu
    \left[
            F_V~\varepsilon^{\mu\nu\alpha\beta}
            p_\alpha k_\beta
            +
            i F_A
            \left(
                g^{\mu\nu} (k p) - p^\mu k^\nu
            \right)
    \right], \\
    M_{T} &=& -i \frac{A_{exp.}}{m f_\pi^{exp.}}~F_T~e^\mu k^\nu~
    \bar u_e(r) \sigma_{\mu \nu} (1+\gamma_5)v_\nu(q) - \nonumber \\
    && -i \frac{A_{exp.}}{m f_\pi^{exp.}}~F'_T~
    (k_\alpha e_\beta - k_\beta e_\alpha)
    \frac{Q^\beta Q_\nu}{Q^2}~
    \bar u_e(r) \sigma^{\alpha \nu} (1+\gamma_5)v_\nu(q),
    \label{Born_amplitude_parts}
\eea

\noindent where $Q=p-k$. The constant $A_{exp.}$ defined in
(\ref{AReplacement}). Squaring amplitude (\ref{BornAmplitude}) and
summing over final photon polarizations leads to the following
decay width:
\bea
    \frac{d^2 \Gamma^0}{dx~dy} &=&\Phi_{tot}^{(0)}(x,y)=
    \frac{M}{(2\pi)^4}A_{exp.}^2 \frac{\pi}{2}
    \left\{
        \Phi_{IB}^{(0)} +
        \left( f_V + f_A \right)^2 \Phi_+^{(0)} +
        \left( f_V - f_A \right)^2 \Phi_-^{(0)}
        + \right.\nonumber \\
        &&\qquad\qquad\qquad+
        2\sqrt{\beta}\left(\left( f_V + f_A \right) \Phi_{Int+}^{(0)} +
        \left( f_V - f_A \right) \Phi_{Int-}^{(0)}\right)
        + \nonumber \\
        &&\qquad\qquad\qquad+
        \left.
        2\left(2 f_T(f_T-f'_T)+f_T'^2\right) \Phi_{T_1}^{(0)}+
        2\left(2 (f_T-f'_T)+f_T' x\right) \Phi_{T_2}^{(0)}
        \frac{}{}
    \right\},   \label{BornDecayWidth}
\eea

\noindent where
\bea
    \Phi_{IB}^{(0)} &=& \frac{(1-y)(1+(1-x)^2)}{x^2(x+y-1)}=
    B(x,y),  \nonumber\\
    \Phi_+^{(0)}    &=& (1-x) (x+y-1)^2,  \nonumber\\
    \Phi_-^{(0)}    &=& (1-x) (1-y)^2, \\
    \Phi_{Int+}^{(0)} &=& -\frac{1}{x}(1-x)(1-y),  \nonumber\\
    \Phi_{Int-}^{(0)} &=&  \frac{1}{x}(1-y)\left(1-x+\frac{x^2}{x+y-1}\right),  \nonumber\\
    \Phi_{T_1}^{(0)} &=& (1-y)(x+y-1), \nonumber \\
    \Phi_{T_2}^{(0)} &=& \frac{1-y}{x},  \nonumber
\label{BornDecayWidthParts} \eea

\noindent where $f_{V,A,T,T'} = (M^2/2mf_\pi)F_{V,A,T,T'} \approx
146 \cdot F_{V,A,T,T'}$. In particular, the conservation of vector
current hypothesis relates the vector formfactor $F_V$ to the
lifetime of the neutral pion
\bea
    |F_V(0)| = \frac{1}{\alpha}
    \sqrt{\frac{2 \Gamma(\pi^0 \to \gamma \gamma)}
           {\pi M}}
    = 0.0259 \pm 0.0005
\eea

\noindent or equivalently  $f_V\approx 3.78$. The Dalitz-plot
distribution of the inner bremsstrahlung part of the Born
amplitude $\Phi_{IB}^{(0)}$ is given in
Table~\ref{BornIBDalitzTable}.

\section{Simplified formula for radiative corrections}
\label{AppendixB}

Here, we present the simplified form of radiative corrections
\bea
    \frac{d^2 \Gamma_{RC}}{dx~dy} &=&
    A_{exp.}^2 \frac{\alpha}{2\pi}
    \left\{
        \Phi_{IB}^{(0)} +
        \left( f_V + f_A \right)^2 \Phi_+^{(0)} +
        \left( f_V - f_A \right)^2 \Phi_-^{(0)}
        + \right.\nonumber \\
        &&\qquad\qquad\qquad+
        2\sqrt{\beta}\left(\left( f_V + f_A \right) \Phi_{Int+}^{(0)} +
        \left( f_V - f_A \right) \Phi_{Int-}^{(0)}\right)
        + \nonumber \\
        &&\qquad\qquad\qquad+
        2\left(2 f_T(f_T-f'_T)+f_T'^2\right) \Phi_{T_1}^{(0)}+
        2\left(2 (f_T-f'_T)+f_T' x\right) \Phi_{T_2}^{(0)}
        \frac{}{}
        + \nonumber \\
        &+&
        \frac{\alpha}{2\pi}\left(L_e-1\right)\left[
        \Phi_{IB}^{(1)} +
        \left( f_V + f_A \right)^2 \Phi_+^{(1)} +
        \left( f_V - f_A \right)^2 \Phi_-^{(1)}
        + \right.\nonumber \\
        &&\qquad\qquad\qquad+
        2\sqrt{\beta}\left(\left( f_V + f_A \right) \Phi_{Int+}^{(1)} +
        \left( f_V - f_A \right) \Phi_{Int-}^{(1)}\right)
        + \nonumber \\
        &&\qquad\qquad+\left.
        \left.
        2\left(2 f_T(f_T-f'_T)+f_T'^2\right) \Phi_{T_1}^{(1)}+
        2\left(2 (f_T-f'_T)+f_T' x\right) \Phi_{T_2}^{(1)}
        \frac{}{}\right]
    \right\},   \label{BornDecayWidthWithRC}
\eea

\noindent where $\Phi_i^{(0)}$ was presented in Appendix
\ref{AppendixA} and
\bea
    \Phi_{IB}^{(1)} &=& \frac{1+\xb^2}{x^2}
                \left[
                    \frac{3}{2}\frac{\yb}{z}+
                    \frac{\yb}{\xb}-
                    \frac{\xb+x y}{\xb^2}\ln y+
                    2\frac{\yb}{z}\ln\frac{\yb}{y}-
                    \frac{x(\xb^2+y^2)}{\xb^2 z}
                    \ln\frac{x}{z}
                \right], \nonumber  \\
    \Phi_+^{(1)}    &=& \xb
                \left[
                    \frac{3}{2}z^2+
                    \frac{1-y^2}{2}+
                    \yb (y-2\xb)+
                    \xb(\xb-2y) \ln y -
                    \xb^2\yb+
                    2z^2 \ln \frac{\yb}{y}
                \right], \nonumber \\
    \Phi_-^{(1)}    &=& \xb
                \left[
                    \frac{3}{2}\yb^2+
                    \frac{1-y^2}{2}+
                    \yb (y-3)+
                    (1-2y) \ln y+
                    2\yb^2 \ln \frac{\yb}{y}
                \right], \\
    \Phi_{Int+}^{(1)} &=& \frac{\xb}{x}
                \left[
                    \frac{\yb}{2} -
                    \yb \ln y -
                    2\yb \ln \frac{\yb}{y}
                \right], \nonumber \\
    \Phi_{Int-}^{(1)} &=& \frac{1}{x}
                \left[
                    -\frac{1}{2}\xb\yb+
                    \frac{3}{2}\frac{x^2\yb}{z}+
                    \xb\left(
                            \yb \ln y +
                            2\yb \ln \frac{\yb}{y}
                       \right) + \right.\nonumber\\
                &&\qquad+\left.
                    x^2\left(
                        \frac{\yb}{\xb} -
                        \frac{\xb+x y}{\xb^2} \ln y +
                        2\frac{\yb}{z} \ln \frac{\yb}{y} -
                        \frac{x(\xb^2+y^2)}{\xb^2 z} \ln
                        \frac{x}{z}
                    \right)
                \right], \nonumber \\
    \Phi_{T_1}^{(1)} &=&
                    \frac{3}{2}\yb z -
                    \frac{1-y^2}{2} +
                    \yb(2\xb+\yb) -
                    (\xb\yb-y) \ln y +
                    2 \yb z \ln \frac{\yb}{y}, \nonumber \\
    \Phi_{T_2}^{(1)} &=& \frac{\yb}{x}
                \left[
                    -\frac{1}{2} +
                    \ln y +
                    2 \ln \frac{\yb}{y}
                \right], \nonumber
\label{BornDecayWidthWithRCParts} \eea

\noindent where $z=x+y-1$,~$\bar x = 1-x$,~$\bar y = 1-y$. Here we
should notice that the functions $\Phi_i^{(1)}$ satisfy the
following property:
\bea
\int\limits_0^1 dy~\Phi_i^{(1)} = 0,
\eea

\noindent which is in accordance with the demands of the
KLN-theorem.

Let us now estimate the magnitude of the terms omitted in our
approximate formula (\ref{BornDecayWidthWithRC}). They are
\bea
    \left(\frac{m}{M}\right)^2 :
    \left|\frac{\alpha}{\pi} K(x,y) \right| :
    \left(\frac{m}{M}\right)^2 \cdot \frac{\alpha}{\pi}
        \ln\frac{M^2}{m^2} :
    \left(\frac{\alpha}{\pi} \ln\frac{M^2}{m^2}\right)^2
    =
    10^{-4} : 1 \div 2 \cdot 10^{-3} :
    10^{-5} : 10^{-4},
\eea

\noindent respectively. The main error arises from
$\frac{\alpha}{\pi}K$ which is presented in (\ref{K_IB}) and
Appendix \ref{AppendixC}. The Dalitz-plot distribution of RC to
the inner bremsstrahlung part of the Born amplitude
$\Phi_{IB}^{(1)}$ is given in Table~\ref{IBRCDalitzTable}.

\section{Hard photon emission $K^{h}_{IB}$-factor}
\label{AppendixC}

In numerical calculation of the $K^{h}_{IB}$-factor
(\ref{Kh_factor}) it is convenient to use the following form of
phase volume (\ref{PhaseVolume}):
$d\Phi_4=\frac{M^4}{(2\pi)^8}\frac{\pi^2}{2^5}\frac{x x_2
y}{\left|A_2\right|}~dx~dy~dC_1~d\Omega_2$, where $d\Omega_2=dC_2
d\phi_2$, $A_2=2-x(1-C_{12})-y(1-C_2)$,
$C_{1,2}=\cos(\widehat{\vec k_{1,2},\vec r})$. Thus the
$K^{h}_{IB}$-factor which comes from hard photon emission RC reads
\bea
    &&K^{h}_{IB}(x,y,\Delta)-2\ln\Delta =
                        \frac{1}{4 \pi}
                        \int
                        dC_1
                        d\Omega_2~
                        \frac{y}{\left|A_2\right|}~\left\{
                           \frac{1}{B(x,y)} I_{NC}(x, x_2, y)
                            +
                            2\frac{x}{x_2}
                        + \right.\nonumber \\
                        &&\left.\qquad\qquad\qquad\qquad\qquad\qquad\qquad\qquad+
                        \frac{1}{B(x,y)}\frac{1}{x'_2}\left(
                             I_{L}(x, x_2, y)
                             -
                             \left|A\right|\frac{a_L}{x_2+y}
                        \right)\right\} + \nonumber \\
                        &&\qquad\qquad\qquad\qquad\qquad\qquad+
                        \int \limits_y^1 \frac{dt}{t}
                        \frac{B(x,t)}{B(x,y)}
                        \left(1 - \frac{y}{t}\right),
                        \label{K_hard}
\eea

\noindent where
\bea I_{L}(x,x_2,y) &=& -y + \frac{2\,y}{a} - y^2 +
\frac{2\,y^2}{a} - \frac{2\,y\,{x}}{a} + \frac{y\,{x}}{b} -
\frac{y\,{x}}{a\,b} -
  \frac{y^2\,{x}}{b} - \frac{2\,y^2\,{x}}{a\,b} +
  \nonumber \\
  &&+
  \frac{y^3\,{x}}{a\,b} + \frac{y\,{{x}}^2}{a\,b} +
  \frac{y^2\,{{x}}^2}{a\,b} + {x_2} - y\,{x_2} - \frac{2\,y\,{x_2}}{a} - \frac{y\,{x_2}}{b} - \frac{2\,y\,{x_2}}{a\,b} +
  \nonumber \\
  &&+
  \frac{y^2\,{x_2}}{b} - \frac{2\,{x_2}}{{x}} + \frac{2\,y\,{x_2}}{{x}} + \frac{4\,y\,{x_2}}{a\,{x}} +
  \frac{{x}\,{x_2}}{b^2} + \frac{{x}\,{x_2}}{b} - \frac{y\,{x}\,{x_2}}{b^2} - \frac{y\,{x}\,{x_2}}{b} +
  \nonumber \\
  &&+
  \frac{2\,y^2\,{x}\,{x_2}}{a\,b} - \frac{{{x}}^2\,{x_2}}{b^2} + \frac{y\,{{x}}^2\,{x_2}}{a\,b} + \frac{{{x_2}}^2}{b} +
  \frac{y\,{{x_2}}^2}{b} - \frac{y\,{x}\,{{x_2}}^2}{b^2} + \frac{y\,{x}\,{{x_2}}^2}{a\,b} -
  \nonumber \\
  &&-
  \frac{{{x}}^2\,{{x_2}}^2}{b^2} -
  \frac{{x}\,{{x_2}}^3}{b^2} + \frac{y}{{{x'}_1}} + \frac{y^3}{{{x'}_1}} - \frac{y\,{x}}{{{x'}_1}} -
  \frac{y\,{x}}{b\,{{x'}_1}} + \frac{y^2\,{x}}{{{x'}_1}} + \frac{2\,y^2\,{x}}{b\,{{x'}_1}} -
  \nonumber \\
  &&-
  \frac{y^3\,{x}}{b\,{{x'}_1}} +
  \frac{y\,{{x}}^2}{b\,{{x'}_1}} - \frac{y^2\,{{x}}^2}{b\,{{x'}_1}} -
  \frac{y\,{x_2}}{{{x'}_1}} - \frac{y\,{x_2}}{b\,{{x'}_1}} + \frac{y^2\,{x_2}}{{{x'}_1}} + \frac{2\,y^2\,{x_2}}{b\,{{x'}_1}} -
  \frac{y^3\,{x_2}}{b\,{{x'}_1}} +
  \nonumber \\
  &&+
  \frac{2\,y\,{x}\,{x_2}}{{{x'}_1}} + \frac{2\,y\,{x}\,{x_2}}{b^2\,{{x'}_1}} +
  \frac{6\,y\,{x}\,{x_2}}{b\,{{x'}_1}} - \frac{4\,y^2\,{x}\,{x_2}}{b^2\,{{x'}_1}} - \frac{6\,y^2\,{x}\,{x_2}}{b\,{{x'}_1}} +
  \nonumber \\
  &&+
  \frac{2\,y^3\,{x}\,{x_2}}{b^2\,{{x'}_1}} - \frac{4\,y\,{{x}}^2\,{x_2}}{b^2\,{{x'}_1}} -
  \frac{4\,y\,{{x}}^2\,{x_2}}{b\,{{x'}_1}} + \frac{4\,y^2\,{{x}}^2\,{x_2}}{b^2\,{{x'}_1}} +
  \frac{2\,y\,{{x}}^3\,{x_2}}{b^2\,{{x'}_1}} +
  \nonumber \\
  &&+
  \frac{y\,{{x_2}}^2}{b\,{{x'}_1}} - \frac{y^2\,{{x_2}}^2}{b\,{{x'}_1}} -
  \frac{4\,y\,{x}\,{{x_2}}^2}{b^2\,{{x'}_1}} - \frac{4\,y\,{x}\,{{x_2}}^2}{b\,{{x'}_1}} +
  \frac{4\,y^2\,{x}\,{{x_2}}^2}{b^2\,{{x'}_1}} +
  \nonumber \\
  &&+
  \frac{4\,y\,{{x}}^2\,{{x_2}}^2}{b^2\,{{x'}_1}} +
  \frac{2\,y\,{x}\,{{x_2}}^3}{b^2\,{{x'}_1}} - \frac{2\,y\,{{x'}_1}}{a} + \frac{y\,{x}\,{{x'}_1}}{a\,b} -
  \frac{y^2\,{x}\,{{x'}_1}}{a\,b} + \frac{2\,{x_2}\,{{x'}_1}}{a} +
  \nonumber \\
  &&+
  \frac{2\,{x_2}\,{{x'}_1}}{b} +
  \frac{2\,{x_2}\,{{x'}_1}}{a\,b} - \frac{2\,{x_2}\,{{x'}_1}}{{x}} - \frac{4\,{x_2}\,{{x'}_1}}{a\,{x}} -
  \frac{{x}\,{x_2}\,{{x'}_1}}{b^2} - \frac{{x}\,{x_2}\,{{x'}_1}}{a\,b} -
  \nonumber \\
  &&-
  \frac{2\,y\,{x}\,{x_2}\,{{x'}_1}}{a\,b} -
  \frac{{x}\,{{x_2}}^2\,{{x'}_1}}{a\,b},
\eea
\bea I_{NC}(x,x_2,y) &=&
\frac{1}{x_2}I_{NC}^{singular}(x,x_2,y)+I_{NC}^{regular}(x,x_2,y),
\eea
\bea I_{NC}^{singular}(x,x_2,y) &=& \frac{-4\,{x}}{a} +
\frac{4\,y\,{x}}{a^2} - \frac{2\,{x}}{{{x'}_1}} +
\frac{2\,y\,{x}}{{{x'}_1}} +
  \frac{4\,y\,{x}}{a\,{{x'}_1}} - \frac{4\,{x}\,{{x'}_1}}{a^2} -
  \nonumber \\
  &&- \frac{4\,{x}\,{{x'}_2}}{a^2} -
  \frac{2\,{x}\,{{x'}_2}}{{{x'}_1}} -
  \frac{4\,{x}\,{{x'}_2}}{a\,{{x'}_1}},\\
I_{NC}^{regular}(x,x_2,y) &=& 2 - \frac{8}{a} + \frac{8\,y}{a^2} -
\frac{4\,y}{a} + \frac{6\,{x}}{a} + \frac{4\,{x}}{a\,b} -
\frac{4\,y\,{x}}{a^2} +
  \frac{3\,y\,{x}}{a\,b} - \frac{y^2\,{x}}{a\,b} - \frac{{{x}}^2}{a\,b} -
  \nonumber \\
  &&-
  \frac{y\,{{x}}^2}{a\,b} + \frac{6\,{x_2}}{a} +
  \frac{4\,{x_2}}{a\,b} - \frac{4\,y\,{x_2}}{a^2} + \frac{3\,y\,{x_2}}{a\,b} - \frac{y^2\,{x_2}}{a\,b} -
  \frac{4\,{x_2}}{a\,{x}} + \frac{4\,y\,{x_2}}{a^2\,{x}} +
  \nonumber \\
  &&+
  \frac{2\,y\,{x}\,{x_2}}{a^2} - \frac{6\,y\,{x}\,{x_2}}{a\,b} -
  \frac{3\,{{x}}^2\,{x_2}}{a\,b} - \frac{{{x_2}}^2}{a\,b} - \frac{y\,{{x_2}}^2}{a\,b} - \frac{3\,{x}\,{{x_2}}^2}{a\,b} -
  \frac{y}{{{x'}_1}} +
  \nonumber \\
  &&+
  \frac{2\,y}{a\,{{x'}_1}} - \frac{y^2}{{{x'}_1}} + \frac{2\,y^2}{a\,{{x'}_1}} + \frac{{x}}{{{x'}_1}} -
  \frac{y\,{x}}{{{x'}_1}} - \frac{2\,y\,{x}}{a\,{{x'}_1}} - \frac{y\,{x}}{b\,{{x'}_1}} -
  \frac{2\,y\,{x}}{a\,b\,{{x'}_1}} +
  \nonumber \\
  &&+
  \frac{y^2\,{x}}{b\,{{x'}_1}} + \frac{{{x}}^2}{b\,{{x'}_1}} +
  \frac{y\,{{x}}^2}{b\,{{x'}_1}} - \frac{2\,y\,{x_2}}{a\,{{x'}_1}} + \frac{y\,{x_2}}{b\,{{x'}_1}} -
  \frac{y\,{x_2}}{a\,b\,{{x'}_1}} - \frac{y^2\,{x_2}}{b\,{{x'}_1}} -
  \nonumber \\
  &&-
  \frac{2\,y^2\,{x_2}}{a\,b\,{{x'}_1}} +
  \frac{y^3\,{x_2}}{a\,b\,{{x'}_1}} + \frac{{x}\,{x_2}}{b^2\,{{x'}_1}} + \frac{{x}\,{x_2}}{b\,{{x'}_1}} -
  \frac{y\,{x}\,{x_2}}{b^2\,{{x'}_1}} - \frac{y\,{x}\,{x_2}}{b\,{{x'}_1}} +
  \nonumber \\
  &&+
  \frac{2\,y^2\,{x}\,{x_2}}{a\,b\,{{x'}_1}} -
  \frac{y\,{{x}}^2\,{x_2}}{b^2\,{{x'}_1}} + \frac{y\,{{x}}^2\,{x_2}}{a\,b\,{{x'}_1}} -
  \frac{{{x}}^3\,{x_2}}{b^2\,{{x'}_1}} + \frac{y\,{{x_2}}^2}{a\,b\,{{x'}_1}} + \frac{y^2\,{{x_2}}^2}{a\,b\,{{x'}_1}} -
  \nonumber \\
  &&-
  \frac{{x}\,{{x_2}}^2}{b^2\,{{x'}_1}} + \frac{y\,{x}\,{{x_2}}^2}{a\,b\,{{x'}_1}} -
  \frac{{{x}}^2\,{{x_2}}^2}{b^2\,{{x'}_1}} - \frac{8\,{{x'}_1}}{a^2} + \frac{4\,{x}\,{{x'}_1}}{a^2} +
  \frac{4\,{x_2}\,{{x'}_1}}{a^2} - \frac{4\,{x_2}\,{{x'}_1}}{a^2\,{x}} -
  \nonumber \\
  &&-
  \frac{2\,{x}\,{x_2}\,{{x'}_1}}{a^2} -
  \frac{8\,{{x'}_2}}{a^2} + \frac{4\,{x}\,{{x'}_2}}{a^2} + \frac{4\,{x_2}\,{{x'}_2}}{a^2} -
  \frac{4\,{x_2}\,{{x'}_2}}{a^2\,{x}} - \frac{2\,{x}\,{x_2}\,{{x'}_2}}{a^2} -
  \nonumber \\
  &&-
  \frac{2\,y\,{{x'}_2}}{a\,{{x'}_1}} +
  \frac{2\,{x}\,{{x'}_2}}{a\,{{x'}_1}} + \frac{2\,{x}\,{{x'}_2}}{b\,{{x'}_1}} + \frac{2\,{x}\,{{x'}_2}}{a\,b\,{{x'}_1}} +
  \frac{y\,{x_2}\,{{x'}_2}}{a\,b\,{{x'}_1}} - \frac{y^2\,{x_2}\,{{x'}_2}}{a\,b\,{{x'}_1}} -
  \nonumber \\
  &&-
  \frac{{x}\,{x_2}\,{{x'}_2}}{b^2\,{{x'}_1}} - \frac{{x}\,{x_2}\,{{x'}_2}}{a\,b\,{{x'}_1}} -
  \frac{2\,y\,{x}\,{x_2}\,{{x'}_2}}{a\,b\,{{x'}_1}} - \frac{{{x}}^2\,{x_2}\,{{x'}_2}}{a\,b\,{{x'}_1}},
\eea
\bea a_L &=& \frac{(1-x_2-y)(1+(1-x)^2)}{x^2(x+x_2+y-1)}\cdot
\frac{(x_2+y)^2+y^2}{x_2+y},\eea

\noindent here $a = \frac{x x_2}{2}(1-C_{12}) - x-x_2$,~~$b =
\frac{x x_2}{2}(1-C_{12})+x'_1+x'_2$,~~ $x'_1=\frac{x_1
y}{2}(1-C_1)$,~~$x'_2=\frac{x_2 y}{2}(1-C_2)$,~~ $C_{12} = C_1
C_2+S_1 S_2 \cos \phi_2$, where $C_{12}=\cos(\widehat{\vec
k_1,\vec k_2})$, $S_{1,2}=\sin(\widehat{\vec k_{1,2},\vec r})$.
$A=x y + x x_2\left(C_2-\frac{C_1 S_2}{S_1}\cos \phi_2\right)$.

Let us note that the combination
$\left(\frac{1}{B(x,y)}\frac{1}{x_2} I_{NC}^{singular}(x, x_2,
y)+2\frac{x}{x_2}\right)$ is finite at $x_2\to0$ limit.
In this integral the value of $x_2$ is fixed by delta-function in
phase volume (\ref{PhaseVolume}):
%
$
    x_2=\frac{1}{A_2}\left(2-2(x+y)+x y (1-C_1)\right)$.
Energy conservation law gives $0 \le x_2 \le 2-x-y$.

\section{Vector and scalar 4-dimensional loop integrals}
\label{AppendixD}

\noindent We introduce the following shorthand for impulse
integrals (we imply real part in r.h.s.):
\begin{gather}
    J_{i j \dots}     = \int \frac{d^4k_1}{i \pi^2}
    \frac{1}{(i)(j)\dots}~,\qquad
    J_{i j \dots}^\mu = \int \frac{d^4k_1}{i \pi^2} \frac{k_1^\mu}{(i)(j)\dots}~,
\end{gather}

\noindent where we have used the short notation for the integral
denominators
\begin{gather}
    (0)  = k_1^2 - \lambda^2, \nonumber\\
    (1)  = (p-k_1)^2 - M^2,   \qquad
    (1') = (p-k-k_1)^2 - M^2, \label{denominators}\\
    (2)  = (r-k_1)^2 - m^2,   \qquad
    (2') = (r+k-k_1)^2 - m^2  \nonumber
\end{gather}

\noindent The integrals with two denominators
\begin{gather}
    J_{1'2}  = L_\Lambda, \qquad
    J_{01}   = L_\Lambda + 1, \qquad
    J_{12}   = L_\Lambda + 1 + \frac{y}{1-y} \ln y, \nonumber \\
    J_{12'}  = L_\Lambda, \qquad
    J_{11'}  = L_\Lambda - 1, \qquad
    J_{01'}  = L_\Lambda + 1 + \frac{x}{1-x} \ln x, \\
    J_{22'}  = L_\Lambda - 1 - L_\beta, \qquad
    J_{02}   = L_\Lambda + 1 - L_\beta, \nonumber \\
    J_{02'}  = L_\Lambda + 1 - \ln(z), \qquad
    J_{1'2'} = L_\Lambda + 1 - \frac{2-y}{1-y} \ln(2-y), \nonumber
\end{gather}

\noindent where: $ 
    L_\Lambda \equiv \ln\frac{\Lambda^2}{M^2},~~
    L_\beta \equiv \ln\beta,
$
~~$\beta = m^2/M^2$.

\noindent The integrals with tree denominators (we put $M=1$ and
introduce the notation $\lambda_0^2=\lambda^2/M^2$)
\begin{eqnarray}
    J_{012}  &=& \frac{1}{2y}\left[L_e~\ln \lambda_0^2 - \ln^2y +
                                    \frac{1}{2}L_\beta^2 + 2 Li_2\left(\frac{y-1}{y}\right)\right], \nonumber \\
    J_{01'2} &=& \frac{1}{1-x}\left[-\xi_2+\ln x~\ln\frac{1-x}{\beta}+ Li_2(x)\right], \nonumber \\
    J_{122'} &=& \frac{1}{1-y}\left\{
                                    \frac{1}{2}\ln^2y + Li_2(1-y)-
                                    \ln \beta \ln y
                              \right\},
\end{eqnarray}
\begin{gather}
    J_{012'} = -\frac{1}{2(1-z)}\ln^2z, \qquad
    J_{011'} = \frac{1}{x}\left[Li_2(1-x)-\xi_2\right], \nonumber \\
    J_{022'} = \frac{1}{z}\left[\frac{1}{2}\ln^2 \frac{z}{\beta}-\xi_2\right],
    \qquad
    J_{11'2} = -\frac{1}{1-y}Li_2(1-y), \nonumber
\end{gather}

\noindent where $L_e = \ln\frac{y^2}{\beta}$, $\xi_2 = Li_2(1) =
\pi^2/6$.

\noindent We also need two integrals with four denominators:
\begin{eqnarray}
    J_{0122'} &=& -\frac{1}{2 y z}
                    \left\{
                        \frac{1}{2} L_e^2
                        +
                        L_e \ln\frac{z^2}{\lambda_0^2} + 2
                        \xi_2
                    \right\},\\
    J_{011'2} &=& \frac{1}{2 x y}
                    \left\{
                        \frac{1}{2} L_e^2
                        +
                        L_e \ln\frac{x^2 \beta}{\lambda_0^2} + 2
                        \xi_2
                    \right\}.
\end{eqnarray}

\noindent Now we consider the vector integrals with tree
denominators:
\begin{gather}
    J_{012}^\mu  = p^\mu~\alpha_{012} + r^\mu~\beta_{012}, \qquad
    J_{01'2}^\mu = (p-k)^\mu~\alpha_{01'2} + r^\mu~\beta_{01'2}, \nonumber \\
    J_{011'}^\mu = p^\mu~\alpha_{011'} + k^\mu~\beta_{011'}, \qquad
    J_{012'}^\mu = p^\mu~\alpha_{012'} + (r+k)^\mu~\beta_{012'}, \nonumber \\
    J_{11'2}^\mu = p^\mu~\alpha_{11'2} + r^\mu~\beta_{11'2} + k^\mu~c_{11'2},
    \qquad
    J_{122'}^\mu = p^\mu~\alpha_{122'} + r^\mu~\beta_{122'} +
    k^\mu~c_{122'},\\
    J_{022'}^\mu = r^\mu~\alpha_{022'} + k^\mu~\beta_{022'}.
    \nonumber
\end{gather}

\noindent The coefficients $\alpha_{ijk}$, $\beta_{ijk}$ and
$c_{ijk}$ are the following:
\begin{eqnarray}
    \alpha_{012}  &=& \frac{1}{y}\left(J_{12}-J_{01}\right), \nonumber \\
    \beta_{012}   &=& \frac{1}{y}\left(J_{12}-J_{02}-\frac{2}{y}\left(J_{12}-J_{01}\right)\right), \nonumber \\
    \alpha_{011'} &=& \frac{1}{x}\left(J_{01}-J_{01'}+x J_{011'}\right), \nonumber \\
    \beta_{011'}  &=& \frac{1}{x}\left(J_{11'}-J_{01'}-\frac{2}{x}\left(J_{01}-J_{01'}+x J_{011'}\right)\right), \nonumber \\
    \alpha_{01'2} &=& \frac{1}{1-x}\left(J_{1'2}-J_{01'}\right), \nonumber \\
    \beta_{01'2}  &=& \frac{1}{1-x}\left(2 J_{01'}-J_{1'2}-J_{02}-x J_{01'2} \right), \nonumber \\
    \alpha_{012'} &=& -\frac{1}{(1-z)^2}\left(2 z \left(J_{12'}-J_{02'}\right)-
                                            (1+z)\left(J_{12'}-J_{01}+z J_{012'}\right)\right), \nonumber \\
    \beta_{012'}  &=& -\frac{1}{(1-z)^2}\left(2 \left(J_{12'}-J_{01}+z J_{012'}\right)-
                                            (1+z)\left(J_{12'}-J_{02'}\right)\right), \\
    \alpha_{022'} &=& \frac{1}{z}\left(J_{02'}-J_{02}+z J_{022'}\right), \nonumber \\
    \beta_{022'}  &=& \frac{1}{z}\left(J_{22'}-J_{02'}\right), \nonumber \\
    \alpha_{11'2} &=& J_{11'2}+\frac{1}{1-y}\left(J_{12}-J_{1'2}\right), \nonumber \\
    \beta_{11'2}  &=& \frac{1}{1-y}\left(J_{1'2}-J_{12}\right), \nonumber \\
    c_{11'2}      &=& \frac{1}{1-y}\left(J_{11'}-2 J_{12}+J_{1'2}-(2-y) J_{11'2}\right), \nonumber \\
    \alpha_{122'} &=& \frac{1}{1-y}\left(J_{12'}-J_{12}\right), \nonumber \\
    \beta_{122'}  &=& J_{122'}+\frac{1}{1-y}\left(J_{12}-J_{12'}\right), \nonumber \\
    c_{122'}      &=& \frac{1}{1-y}\left(2 J_{12}-J_{12'}-J_{22'}-y J_{122'}\right). \nonumber
\end{eqnarray}

\noindent The vector integrals with four denominators:
\begin{eqnarray}
    J_{0122'}^\mu &=& p^\mu~\alpha_{0122'} + r^\mu~\beta_{0122'} + (r+k)^\mu~c_{0122'}.
\end{eqnarray}

\noindent The coefficients $\alpha_{ijk}$, $\beta_{ijk}$ and
$c_{ijk}$ are the following:
\begin{eqnarray}
    \alpha_{0122'} &=&
    \frac{1}{2(1-x)}\left[J_{122'}+\frac{1}{1-y}\left(z J_{022'}-(1-2y+z)J_{012'}-
                                                    y\left(J_{012}-z J_{0122'}\right)\right)
                                            \right], \nonumber \\
    \beta_{0122'}  &=& \frac{1}{2(1-x)(1-y)z}\left[
                            (z+y-z y-1)J_{122'} -
                            z(1-2y+z)J_{022'}+(1-z)^2J_{012'}+
                            \right. \nonumber \\
                            &&\quad\quad\quad\quad\quad\quad\quad\quad\left.+
                            (y(1+z)-2z)\left(z J_{0122'}-J_{012}\right)
                                            \right], \\
    c_{0122'}      &=& \frac{1}{2z(1-x)}\left[
                                (y-2z)J_{122'}+
                                \right.\nonumber \\
                                &&\quad\quad\quad\quad\quad+\left.
                                \frac{1}{1-y}
                                \left(
                                    y^2\left(J_{012}-z J_{0122'}\right)
                                    -y z J_{022'}-
                                    (y(1+z)-2z)J_{012'}
                                \right)
                            \right]. \nonumber
\end{eqnarray}

\begin{table}[pp]
\begin{tabular}{|c|r|r|r|r|r|r|r|r|r|}
\hline
y / x & 0.2 & 0.3 & 0.4 & 0.5 & 0.6 & 0.7 & 0.8 & 0.9 \\
\hline
0.9 & 41.000 & 8.278 & 2.833 & 1.250 & 0.644 & 0.371 & 0.232 & 0.156 \\
0.8 & & 33.111 & 8.500 & 3.333 & 1.611 & 0.889 & 0.542 & 0.356 \\
0.7 & & & 25.500 & 7.500 & 3.222 & 1.668 & 0.975 & 0.623 \\
0.6 & & & & 20.000 & 6.444 & 2.966 & 1.625 & 0.998 \\
0.5 & & & & & 16.111 & 5.561 & 2.708 & 1.558 \\
0.4 & & & & & & 13.347 & 4.875 & 2.494 \\
0.3 & & & & & & & 11.375 & 4.364 \\
0.2 & & & & & & & & 9.975 \\
\hline
\end{tabular}
\caption{The value of $\Phi_{IB}^{(0)}$ (i.e. Born inner
bremsstrahlung part, see (\ref{BornDecayWidth})).}
\label{BornIBDalitzTable}

\begin{tabular}{|c|r|r|r|r|r|r|r|r|r|}
\hline
y / x & 0.2 & 0.3 & 0.4 & 0.5 & 0.6 & 0.7 & 0.8 & 0.9 \\
\hline
0.9 & -2.740 & -0.521 & -0.174 & -0.076 & -0.039 & -0.022 & -0.014 & -0.009 \\
0.8 & & -1.773 & -0.407 & -0.152 & -0.071 & -0.039 & -0.023 & -0.015 \\
0.7 & & & -1.155 & -0.290 & -0.116 & -0.057 & -0.032 & -0.020 \\
0.6 & & & & -0.772 & -0.203 & -0.084 & -0.043 & -0.025 \\
0.5 & & & & & -0.521 & -0.140 & -0.059 & -0.031 \\
0.4 & & & & & & -0.348 & -0.093 & -0.040 \\
0.3 & & & & & & & -0.221 & -0.056 \\
0.2 & & & & & & & & -0.118 \\
\hline
\end{tabular}
\caption{The value of
$\frac{\alpha}{2\pi}\left(L_e-1\right)\Phi_{IB}^{(1)}$ (see
(\ref{BornDecayWidthWithRC})).} \label{IBRCDalitzTable}

\begin{tabular}{|c|r|r|r|r|r|r|r|r|r|}
\hline
y / x & 0.2 & 0.3 & 0.4 & 0.5 & 0.6 & 0.7 & 0.8 & 0.9 \\
\hline 0.9 & -2.568 & -1.855 & -1.135 & -0.512 & -0.160 & 0.106 & 0.215 & -0.018 \\
0.8 & & -2.707 & -2.362 & -1.978 & -1.596 & -1.229 & -0.994 & -1.161 \\
0.7 & & & -2.657 & -2.493 & -2.248 & -2.012 & -1.850 & -1.941 \\
0.6 & & & & -2.716 & -2.600 & -2.438 & -2.333 & -2.431 \\
0.5 & & & & & -2.892 & -2.748 & -2.661 & -2.768 \\
0.4 & & & & & & -3.137 & -2.979 & -3.049 \\
0.3 & & & & & & & -3.575 & -3.405 \\
0.2 & & & & & & & & -4.324 \\
\hline
\end{tabular}
\caption{The value of $K(x,y)$ (see (\ref{RCResultTotal})).}
\label{KDalitzTable}
\end{table}



\end{document}